\documentclass[twocolumn,showpacs,showkeys,preprintnumbers]{revtex4-1}
\usepackage{amssymb}

\usepackage{amsfonts}

\usepackage{latexsym}

\usepackage{dcolumn}

\usepackage{amsmath}

\usepackage{graphicx}

\usepackage{psfrag,pstricks}

\begin{document}

\title{Squeezed-state generation via nonlinear atom-atom interaction in the atomic field of a Bose-Einstein condensate interacting with an optical cavity }

\author{A. Dalafi$^{1}$ }
\email{adalafi@yahoo.co.uk}

\author{M. H. Naderi$^{1,2}$}
\author{M. Soltanolkotabi$^{1,2}$}

\affiliation{$^{1}$ Department of Physics, Faculty of Science, University of Isfahan, Hezar Jerib, 81746-73441, Isfahan, Iran\\
$^{2}$Quantum Optics Group, Department of Physics, Faculty of Science, University of Isfahan, Hezar Jerib, 81746-73441, Isfahan, Iran}

\date{\today}

\begin{abstract}
In this paper, we investigate theoretically a system consisting of a one dimensional Bose-Einstein condensate trapped inside the optical lattice of an optical cavity. In the weak-interaction regime and under the Bogoliubov approximation, the wave function of the Bose-Einstein condensate can be described by a classical field (condensate mode) having some quantum fluctuations (the Bogoliubov mode) about the mean value. Such a system behaves as a so-called atomic parametric amplifier, similar to an optical parametric amplifier, where the condensate and the Bogoliubov modes play respectively, the roles of the pump field and the signal mode in the degenerate parametric amplifier and the \textit{s}-wave scattering frequency of atom-atom interaction plays the role of the nonlinear gain parameter . We show that using the nonlinear effect of atomic collisions, how one can manipulate and control the state of the Bogoliubov mode and produce squeezed states.
\end{abstract}

\pacs{42.50.Dv, 42.50.Wk, 67.85.Hj} 
\keywords {Nonclassical-state generation, Bose-Einstein condensate, atomic collisions, optomechanical coupling}
\maketitle

\section{Introduction}
The radiation pressure coupling between the optical field inside an optomechanical cavity and its moving mirror has been employed for a wide range of applications such as the cooling of the vibrational mode of the moving mirror to its quantum mechanical ground state \cite{Genes 2008, Bhattacharya}, ultrahigh precision measurements \cite{LaHaye} and the detection of gravitational waves \cite{Abramovici} and also providing a good approach for fundamental studies of the transition between the quantum and the classical world\cite{Bradaschia, Kippenberg,marshall}.

On the other hand, in hybrid systems consisting of a Bose-Einstein condensate (BEC) trapped inside a high finesse optical cavity \cite{Brenn Nature,Gupta,Brenn Science} interacting dispersively with the optical field of the cavity, an effective optomechanical coupling comes into existence in which the fluctuations of the atomic field of the BEC (the Bogoliubov mode) plays the role of the vibration mode of the moving mirror in an optomecanocal cavity. For low photon numbers or in the weakly interacting regime the dynamics can be restricted to the first motional mode of the BEC which plays the role of the mechanical oscillator\cite{Kanamoto 2010, Nagy Ritsch 2009}.  

Optomechanical systems have also attracted considerable attention in connection with quantum state engineering; because of the great possibilities they are expected to produce nonclassical states of both the mechanical oscillator \cite{cohadon} and the cavity field \cite{mancini02}. From the point of view of quantum mechanics a system which  may be a macroscopic object like the moving mirror of an optomechanical cavity can be in a coherent superposition of different quantum states. Recently, due to improved technology, there has been a growing interest in the possibility of observing such superposition states, commonly known as Schr\"{o}dinger cat states \cite{bose97}. Coherent states of the electromagnetic field mode inside an opical cavity is a good candidate for these macroscopic states \cite{mancini97}. Recently, it has been shown the possibility of generating motional nonlinear coherent states and their superposition for an undamped vibrating micromechanical membrane inside an optical cavity \cite{barzanjehgeneration}.

One of the most important characteristics of the optomechanical systems is a kind of inherent nonlinearity which is due to the mutual interaction between the optical field and the matter inside (the moving mirror or the atomic ensemble) \cite{Gong,meys, McCullen,Meiser}. This nonlinearity leads to realization of the Kerr effect in such systems \cite{Gong}. In hybrid optomechanical systems containing a BEC there exists another kind of nonlinearity which is due to the atom-atom interaction. Both kinds of these nonlinearities have considerable effects on the optical properties of the system like the bistability of the cavity \cite{dalafi2} and the squeezing of the output optical field \cite{Bhattacherjee10} and also on the mechanical properties like the cooling process of the moving mirror \cite{Barzanjeh2}.

In a previous paper \cite{dalafi1} we showed that the nonlinear atom-atom interaction in a hybrid system consisting of a BEC inside an optical cavity is very similar to the interaction Hamiltonian of a degenerate parametric amplifier (DPA) which can lead to the normal mode splitting (NMS) phenomenon. In a DPA a pump beam generates a signal beam by interacting with a $ \chi^{(2)} $ nonlinearity. This process has long been considered as an important source of the squeezed state of the radiation field \cite{zubairybook}.

In the present paper, we consider a one dimensional BEC interacting dispersively with the optical field of an optical cavity. In the weak-interaction regime and using the Bogoliubov approximation, the BEC can be described by a single mode quantum field (the Bogoliubov mode) which fluctuates about a classical mean field (condensate mode). Such a system behaves as a so-called atomic parametric amplifier (APA) in which the condensate acts as an atomic pump field and the Bogoliubov mode plays the role of the signal mode in the DPA. Besides, the \textit{s}-wave scattering frequency of atom-atom interaction plays the role of the nonlinear gain parameter. 

In the absence of damping processes, we calculate the time evolution of the state vector of the system and show that the degree of squeezing of the quadratures of the Bogoliubov mode can be controlled by the \textit{s}-wave scattering frequency of atomic collisions. Since the \textit{s}-wave scattering frequency is controllable through the transverse trapping potential \cite{Morsch} then the degree of squeezing of the Bogoliubov mode becomes controllable.

The paper is structured as follows. In section \ref{secH} we derive the Hamiltonian of the system and diagonalize it in two steps. In section \ref{secDyn} the evolution operator of the system in the Schr\"{o}dinger picture and in the absence of damping processes is calculated and then the reduced density operator of the Bogoliubov mode of the BEC is derived. In section \ref{secQ} the effect of atomic collisions on the $ Q $ function of the Bogoliubov mode is investigated. In section \ref{secQuad} we will show how one can manipulate the suqeezing degree of the quadratures of the Bogoliubov mode throught the \textit{s}-wave scattering frequency. Finally, our conclusions are
summarized in section \ref{secCon}.

\section{System Hamiltonian}\label{secH}
\begin{figure}[ht]
\centering
\includegraphics[width=2in]{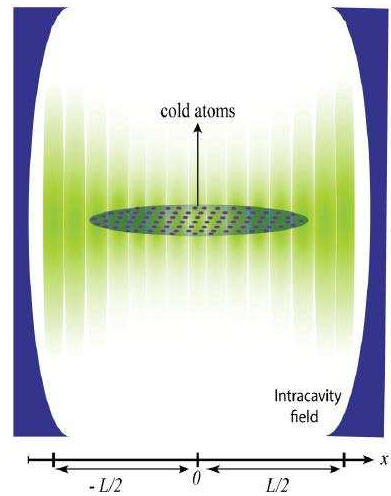}
\caption{
(Color online) N two-level atoms trapped in an optical cavity interacting dispersively with a single cavity mode.}
\label{fig:fig1}
\end{figure}

As schematically shown in Fig.\ref{fig:fig1}, we consider a system consisting of a BEC of $N$ two-level atoms with mass $m_{0}$ and transition frequency $\omega_{a}$ inside an optical cavity with length $L$. The cavity is driven through one of its mirrors by a laser with frequency $\omega_{p}$, and wavenumber $k=\omega_{p}/c$. We assume the BEC to be confined in a cylindrically symmetric trap with a transverse trapping frequency $\omega_{\mathrm{\perp}}$ and negligible longitudinal confinement along the $x$ direction \cite{Morsch}. In this way we can describe the dynamics within an effective one-dimensional model by quantizing the atomic motional degree of freedom along the $x$ axis only.

In the dispersive regime where the laser pump is far detuned from the atomic resonance ($\Delta_{a}=\omega_{p}-\omega_{a}$  exceeds the atomic linewidth $\gamma$ by orders of magnitude), the excited electronic state of the atoms can be adiabatically eliminated and spontaneous emission can be neglected \cite{Masch Ritch 2004}. In the frame rotating at the pump frequency, the many-body Hamiltonian reads
\begin{eqnarray}\label{H1}
H&=&-\hbar\Delta_{c} a^{\dagger} a + i\hbar\eta  (a^{\dagger}-a)+\int_{-L/2}^{L/2} dx \Psi^{\dagger}(x)\Big[\frac{-\hbar^{2}}{2m_{0}}\frac{d^{2}}{dx^{2}}\nonumber\\
&&+\hbar U_{0} \cos^2(kx) a^{\dagger} a+\frac{1}{2} U_{s}\Psi^{\dagger}(x)\Psi(x)\Big] \Psi(x).
\end{eqnarray}
Here, $a$ is the annihilation operator of the optical field, $\Delta_{c}=\omega_{p}-\omega_{c}$ is the cavity-pump detuning, $U_{0}=g_{0}^{2}/\Delta_{a}$ is the optical lattice barrier height per photon which represents the atomic back action on the field, $g_{0}$ is the vacuum Rabi frequency, $U_{s}=\frac{4\pi\hbar^{2} a_{s}}{m_{0}}$ and $a_{s}$ is the two-body \textit{s}-wave scattering length \cite{Masch Ritch 2004,Dom JB}. 

In the weakly interacting regime, where $U_{0}\langle a^{\dagger}a\rangle\leq 10\omega_{R}$ ($\omega_{R}=\frac{\hslash k^{2}}{2m_{0}}$ is the recoil frequency of the condensate atoms), and under the Bogoliubov approximation \cite{Nagy Ritsch 2009}, the atomic field operator can be expanded as
\begin{equation}\label{opaf}
\Psi(x)=\sqrt{\frac{N}{L}}+\sqrt{\frac{2}{L}}\cos(2kx) c,
\end{equation}
where the first term is the condensate mode which is considered as a c-number and the operator $c$ in the second term is the annihilation operator of the Bogoliubov mode. Substituting this expansion into Eq.(\ref{H1}), we can find the Hamiltonian of the system in the following form
\begin{equation}\label{Hopa}
H=\hbar\delta_{c}a^{\dagger}a+\hbar\Omega_{c}c^{\dagger}c+\frac{1}{4}\hbar\omega_{sw}(c^{2}+c^{\dagger2})+\frac{\sqrt{2}}{2}\hbar\zeta a^{\dagger}a (c+c^{\dagger}).
\end{equation}
 Here, we have assumed that the external laser drives the cavity for a limited time until the optical field and the Bogoliubov mode are prepared in a coherent and a vacuum state, respectively and then it is turned off and we let the system evolves by itself. In Eq.(\ref{Hopa}) $\delta_{c}=-\Delta_{c}+\frac{1}{2}N U_{0}$ is the cavity effective detuning, $ \Omega_{c}=4\omega_{R}+\omega_{sw} $ the frequency of the Bogoliubov mode, $ \omega_{sw}=8\pi\hbar a_{s}N/m_{0}Lw^2 $ the s-wave scattering frequency and $w$ is the waist of the optical potential.

The third term of the Hamiltonian (\ref{Hopa}) corresponds to the atom-atom interaction and the fourth term is the optomechanical interaction of the Bogoliubov mode with the radiation pressure of the optical field with the coupling constant $ \zeta=\frac{1}{2}\sqrt{N}U_{0} $. 

In a DPA due to the nonlinear interaction of the pump field with a nonlinear crystal a signal mode in a sqeezed state is generated \cite{zubairybook}. The interaction Hamiltonian of such a system is
\begin{equation}
\mathcal{V}=\hbar\mathcal{G}(a^{\dagger2}b+a^2b^{\dagger}),
\end{equation}
where the operators $ b $ and $ a $ are the annihilation operators of the driving field and the signal mode, respectively and $ \mathcal{G} $ is the coupling constant which depends on the second-order susceptibility tensor that mediates the interaction. In the limit of strong driving field when its depletion is negligible, it can be considered as a classical field and therefore the interaction Hamiltonian can be written as
\begin{equation}\label{HDPA}
\mathcal{V}=\hbar\mathcal{G}\beta_{p}(a^{\dagger2}e^{-i\phi}+a^{2}e^{i\phi}),
\end{equation}
where $ \beta_{p} $ and $ \phi $ are, respectively, the amplitude and the phase of the driving field.

As is seen, the interaction Hamiltonian of Eq.(\ref{HDPA}) is very similar to the atom-atom interaction in the Hamiltonian (\ref{Hopa}) where the Bogoliubov mode $ c $ (with zero phase) plays the role of the signal mode $ a $. Besides, the \textit{s}-wave scattering frequency ($ \omega_{sw} $) plays the role of $ \mathcal{G}\beta_{p} $ in the DPA Hamiltonian. In this way, we are dealing with an APA where the condensate mode acts as the driving field and the Bogoliubov mode acts as the signal mode. 

In the following we will diagonalize the Hamiltonian (\ref{Hopa}) in two steps and will show that the atom-atom interaction term is responsible for the generation of squeezed states in the Bogoliubov mode.

\subsection{The first step of diagonalization}
Since the radiation pressure of the optical field on the Bogoliubov mode acts as a driving field, so in the first step of diagonalization we use the displacement operator
\begin{equation}\label{D}
D(\beta a^{\dagger}a)=\exp\Big(\beta a^{\dagger}a(c^{\dagger}-c)\Big),
\end{equation}
where the parameter $ \beta $ is defined as
\begin{equation}
\beta=\frac{\sqrt{2}}{2}\frac{\zeta}{\Omega_{c}+\frac{1}{2}\omega_{sw}}.
\end{equation}
The action of this displacement unitary transformation on the operators $ c $ and $ c^{\dagger} $ will give the following results:
\begin{subequations}\label{DcD}
\begin{eqnarray}
c^{\prime}&=&D(\beta a^{\dagger}a) c D^{\dagger}(\beta a^{\dagger}a)= c-\beta a^{\dagger}a,\label{DcDa}\\
c^{\prime\dagger}&=&D(\beta a^{\dagger}a) c^{\dagger} D^{\dagger}(\beta a^{\dagger}a)= c^{\dagger}-\beta a^{\dagger}a\label{DcDb}.
\end{eqnarray}
\end{subequations}
Transforming the Hamiltonian (\ref{Hopa}) under the unitary operator (\ref{D}), one can find
\begin{eqnarray}\label{DHD1}
H^{\prime}&=&D(\beta a^{\dagger}a) H D^{\dagger}(\beta a^{\dagger}a),\nonumber\\
&&=\hbar\delta_{c}a^{\dagger}a+\hbar\Omega_{c}c^{\prime\dagger}c^{\prime}\nonumber\\
&&+\frac{1}{4}\hbar\omega_{sw}(c^{\prime2}+c^{\prime\dagger2})+\frac{\sqrt{2}}{2}\hbar\zeta a^{\dagger}a (c^{\prime}+c^{\prime\dagger}).
\end{eqnarray}
After substituting Eqs.(\ref{DcDa}, \ref{DcDb}) for $ c^{\prime} $ and $ c^{\prime\dagger} $ in Eq. (\ref{DHD1}) the Hamiltonian $ H^{\prime} $ reads
\begin{eqnarray}\label{Hprime}
H^{\prime}&=&\hbar\delta_{c}a^{\dagger}a-\hbar(\Omega_{c}+\frac{1}{2}\omega_{sw})\beta^2(a^{\dagger}a)^2+\hbar\Omega_{c}c^{\dagger}c\nonumber\\
&&+\frac{1}{4}\hbar\omega_{sw}(c^{2}+c^{\dagger2}).
\end{eqnarray}
As is seen from this equation, the energy structure of the cavity-field Hamiltonian, corresponding to the first two terms in the right hand side of Eq.(\ref{Hprime}), becomes the anharmonic one due to the photon-photon interaction induced by the radiation pressure. The Hamiltonian (\ref{Hprime}) is diagonal in terms of the creation and annihilation operators of the optical field, while it is not diagonal in terms of the operators of the Bogoliubov mode due to the presence of atom-atom interaction term.

\subsection{The second step of diagonalization}
In order to diagonalize the Hamiltonian (\ref{Hprime}) in terms of the operators of the Bogoliubov mode, $ c $ and $ c^{\dagger} $, one can use the squeezing operator
\begin{equation}\label{S}
S(\xi)=\exp\Big(\frac{1}{2}\xi(c^{2}-c^{\dagger2})\Big),
\end{equation}
where $ \xi $ is the squeezing parameter. Transforming the operators $ c $ and $ c^{\dagger} $ under this unitary transformation leads to the following operators:
\begin{subequations}\label{ScS}
\begin{eqnarray}
c^{\prime\prime}&=&S(\xi) c S^{\dagger}(\xi)= \mu c+\nu c^{\dagger},\label{ScSa}\\
c^{\prime\prime\dagger}&=&S(\xi) c^{\dagger} S^{\dagger}(\xi)= \nu c+\mu c^{\dagger},\label{ScSb}
\end{eqnarray}
\end{subequations}
where $ \mu=\cosh\xi $ and $ \nu=\sinh\xi $ are defined in terms of the squeezing parameter $ \xi $.

The squeezing parameter should be so evaluated that the squeezing operator $ S(\xi) $ diagonalizes the Hamiltonian (\ref{Hprime}). If the unitary transformation $ S(\xi) $ is applied to Eq.(\ref{Hprime}), then the Hamiltonian in Eq.(\ref{Hprime}) becomes
\begin{eqnarray}\label{SHS1}
H^{\prime\prime}&=&S(\xi) H^{\prime} S^{\dagger}(\xi),\nonumber\\
&&=\hbar\delta_{c}a^{\dagger}a-\hbar(\Omega_{c}+\frac{1}{2}\omega_{sw})\beta^2(a^{\dagger}a)^2\nonumber\\
&&+\hbar\Omega_{c}c^{\prime\prime\dagger}c^{\prime\prime}+\frac{1}{4}\hbar\omega_{sw}(c^{\prime\prime2}+c^{\prime\prime\dagger2}).
\end{eqnarray}
By using Eqs.(\ref{ScSa},\ref{ScSb}), the Hamiltonian of Eq.(\ref{SHS1}) takes the following form
\begin{eqnarray}\label{SHS2}
H^{\prime\prime}&=&\hbar\Big[\Omega_{c}(\mu^2+\nu^2)+\mu\nu\omega_{sw})\Big] c^{\dagger}c\nonumber\\
&&+\hbar\Big[\mu\nu\Omega_{c}+\frac{1}{4}\omega_{sw}(\mu^2+\nu^2)\Big](c^{2}+c^{\dagger2})\nonumber\\
&&+\hbar\delta_{c}a^{\dagger}a-\hbar(\Omega_{c}+\frac{1}{2}\omega_{sw})\beta^2(a^{\dagger}a)^2+\mathcal{E}_{0}.
\end{eqnarray}
In this Hamiltonian $ \mathcal{E}_{0}=\hbar\Omega_{c}\nu^{2}+\frac{1}{2}\hbar\omega_{sw}\mu\nu $ is a constant which has no effect on the dynamics of the system. On the other hand, the term in the square brackets of the first term of the Hamiltonian (\ref{SHS2}) is the effective frequency of the Bogoliubov mode which we will denote it by the new parameter $ \Omega^{\prime}_{c} $. Furthermore, in order to bring the Hamiltonian (\ref{SHS2}) into the diagonal form, the term in the square brackets in the second line of Eq.(\ref{SHS2}) should be equal to zero. In this way, we will have the following system of algebraic equations:
\begin{subequations}
\begin{eqnarray}\label{sys4}
\Omega_{c}(\mu^2+\nu^2)+\omega_{sw} \mu\nu&=&\Omega^{\prime}_{c},\\
\Omega_{c} \mu\nu+\frac{1}{4}\omega_{sw} (\mu^2+\nu^2)&=&0,\\
\mu^2-\nu^2=1,
\end{eqnarray}
\end{subequations}
With this system of algebraic equations the Hamiltonian $ H^{\prime\prime} $ takes a diagonal form in terms of the operators $ c $ and $ c^{\dagger} $ as follows:
\begin{equation}\label{Hzegond}
H^{\prime\prime}=\hbar\Omega^{\prime}_{c}c^{\dagger}c+\hbar\delta_{c}a^{\dagger}a-\hbar(\Omega_{c}+\frac{1}{2}\omega_{sw})\beta^2(a^{\dagger}a)^2.
\end{equation}

Solving the system of algebraic equations for the parameters $ \mu $ and $ \nu $, they are obtained as
\begin{subequations}\label{munu}
\begin{eqnarray}
\mu&=&\frac{1}{\sqrt{2}}\sqrt{\frac{\Omega_{c}}{\Omega^{\prime}_{c}}+1},\label{munua}\\
\nu&=&\frac{1}{\sqrt{2}}\sqrt{\frac{\Omega_{c}}{\Omega^{\prime}_{c}}-1}\label{munub},
\end{eqnarray}
\end{subequations}
where the effective frequency of the Bogoliubov mode $ \Omega^{\prime}_{c} $ is
\begin{eqnarray}\label{Omepc}
\Omega_{c}^{\prime}&=&\sqrt{\Omega_{c}^{2}-\frac{1}{4}\omega_{sw}^{2}},\nonumber\\
&&=\sqrt{\Big(4\omega_{R}+\frac{1}{2}\omega_{sw}\Big)\Big(4\omega_{R}+\frac{3}{2}\omega_{sw}\Big)}.
\end{eqnarray}
In this way using Eqs.(\ref{munua}, \ref{munub}), one can obtain the squeezing parameter $ \xi=\cosh^{-1}\mu=\sinh^{-1}\nu $.

Therefore using the two unitary transformations, i.e., the displacement and the squeezing operators [Eqs.(\ref{D}), (\ref{S})], and based on Eqs.(\ref{SHS1}),(\ref{DHD1}) the Hamiltonian of the system can be transformed into the diagonalized Hamiltonian:
\begin{equation}\label{HzegondH}
H^{\prime\prime}=S(\xi)D(\beta a^{\dagger}a) H D^{\dagger}(\beta a^{\dagger}a)S^{\dagger}(\xi).
\end{equation}

\section{Dynamics of the System}\label{secDyn}
In this section we are going to study the time evolution of the state vector of the system in the Schr\"{o}dinger picture in the absence of any damping processes i.e., the time interval in which we observe the system is smaller than the decoherence times of both the optical and the atomic fields.

In order to obtain the time evolution of the state vector of the system in the Schr\"{o}dinger picture we need to calculate the time evolution operator $ U(t)=\exp\Big(-\frac{it}{\hbar}H\Big) $. However, since the Hamiltonian $ H $ is not diagonal we use Eq.(\ref{HzegondH}) to writ it in terms of the diagonalized Hamiltonian $ H^{\prime\prime} $ in the following form:
\begin{equation}\label{HHzegond}
H=D^{\dagger}(\beta a^{\dagger}a)S^{\dagger}(\xi) H^{\prime\prime} S(\xi)D(\beta a^{\dagger}a).
\end{equation}
Defining the unitary operator $ X= S(\xi)D(\beta a^{\dagger}a) $ and using Eq.(\ref{HHzegond}) the time evolution operator can be written as follows:
\begin{eqnarray}\label{U}
U(t)&=&\exp\Big(-\frac{it}{\hbar}X^{\dagger}H^{\prime\prime}X\Big),\nonumber\\
&&=X^{\dagger}\exp\Big(-\frac{it}{\hbar}H^{\prime\prime}\Big)X,\nonumber\\
&&=D^{\dagger}(\beta a^{\dagger}a)S^{\dagger}(\xi)\exp\Big(-\frac{it}{\hbar}H^{\prime\prime}\Big)S(\xi)D(\beta a^{\dagger}a).\nonumber\\
\end{eqnarray}
In the second line we have used the operator theorems \cite{Louisell}. Since $  H^{\prime\prime} $ is diagonal, calculation of the exponential function appeared in the last line of Eq.(\ref{U}) is straightforward.

\subsection{Evolution of the state vector of the system}
Now using the time evolution operator in the form obtained in Eq.(\ref{U}) we can investigate the time evolution of the state vector of the system. If we assume the optical field of the cavity and the Bogoliubov mode of the BEC have been initially prepared, respectively, in a coherent and a vacuum state, then the initial state vector of the system is $ |\psi(0)\rangle=|\alpha\rangle_{a}\otimes|0\rangle_{c} $ where the indices $ a $ and $ c $ refer to the optical field of the cavity and the Bogoliubov mode of the condensate, respectively.

In a previous paper \cite{dalafi2}, we showed that by controlling the detuning between the frequencies of the laser pump and the cavity resonance, the fluctuation in the number of atoms in the Bogoliubov mode can be minimized. If it reduces to values below unity, one can conclude that the Bogoliubov mode has been prepared in the vacuum state.

Therefore, the time evolution of the state vector of the system can be evaluated as follows:
\begin{eqnarray}
|\psi(t)\rangle&=&U(t)|\psi(0)\rangle,\nonumber\\
&&=D^{\dagger}(\beta a^{\dagger}a)S^{\dagger}(\xi) e^{-\frac{it}{\hbar}H^{\prime\prime}}S(\xi)D(\beta a^{\dagger}a)|\psi(0)\rangle.\nonumber\\
\end{eqnarray}
By expanding the coherent state of the optical field in terms of number states, the state vector of the system can be written as
\begin{eqnarray}
|\psi(t)\rangle&=&e^{-\frac{1}{2}|\alpha|^2}\sum_{n=0}^{\infty}\frac{\alpha^n}{\sqrt{n!}}D^{\dagger}(\beta a^{\dagger}a)S^{\dagger}(\xi) \nonumber\\
&&\times e^{-\frac{it}{\hbar}H^{\prime\prime}} |n\rangle_{a}\otimes |\beta n, \xi\rangle_{c},
\end{eqnarray}
where $ |\beta n, \xi\rangle_{c}=S(\xi)D(\beta n)|0\rangle_{c} $ is a squeezed coherent state of the Bogoliobov mode ($ c $) of the condensate. By substituting the right hand side of Eq.(\ref{Hzegond}) for $ H^{\prime\prime} $ in the above equation we will have:
\begin{eqnarray}
|\psi(t)\rangle&=&e^{-\frac{1}{2}|\alpha|^2}\sum_{n=0}^{\infty}\frac{\alpha^n}{\sqrt{n!}}e^{-in\delta_{c}t}e^{i\beta^2n^2(\Omega_{c}+\frac{1}{2}\omega_{sw})t}|n\rangle_{a}\nonumber\\
&&\otimes D^{\dagger}(\beta n)S^{\dagger}(\xi)e^{-i\Omega^{\prime}_{c}t c^{\dagger}c}|\beta n, \xi\rangle_{c}.
\end{eqnarray}
Here, we have used the eigenvalue equation $ a^{\dagger}a|n\rangle_{a}=n|n\rangle_{a} $. The last expression in this equation, i.e., $ e^{-i\Omega^{\prime}_{c}t c^{\dagger}c}|\beta n, \xi\rangle_{c} $ denotes the free evolution of a squeezed coherent sate which is given by [Appendix \ref{apA}]:
\begin{equation}
e^{-i\Omega^{\prime}_{c}t c^{\dagger}c}|\beta n, \xi\rangle_{c}=|\beta ne^{-i\Omega^{\prime}_{c}t}, \xi e^{-i2\Omega^{\prime}_{c}t}\rangle_{c}.
\end{equation}
Using this equation, the state vector of the system at time $ t $ reads
\begin{eqnarray}\label{psitotal}
|\psi(t)\rangle &=& e^{-\frac{1}{2}|\alpha|^2}\sum_{n=0}^{\infty}\frac{\alpha^n}{\sqrt{n!}}e^{-in\delta_{c}t} e^{i\beta^2n^2(\Omega_{c}+\frac{1}{2}\omega_{sw})t}|n\rangle_{a},\nonumber\\
&&\otimes D^{\dagger}(\beta n)S^{\dagger}(\xi)|\beta ne^{-i\Omega^{\prime}_{c}t}, \xi e^{-i2\Omega^{\prime}_{c}t}\rangle_{c}.
\end{eqnarray}

\subsection{The reduced density operator of the Bogoliubov mode}
In order to determine the reduced density operator of the Bogoliubov mode, we first construct the density operator of the total system $ \rho(t)=|\psi(t)\rangle\langle\psi(t)| $ where $ |\psi(t)\rangle $ is given by Eq.(\ref{psitotal}). Then, taking trace over the degrees of freedom of the optical field, i.e., $ \rho_{c}(t)=tr_{a}[\rho(t)] $, the reduced density operator of the Bogoliubov mode is obtained as follows:
\begin{equation}\label{rhoc}
\rho_{c}(\tau)=e^{-|\alpha|^2}\sum_{n=0}^{\infty}\frac{|\alpha|^{2n}}{n!}|\phi_{n,\xi}(\tau)\rangle\langle\phi_{n,\xi}(\tau)|,
\end{equation}
where the state vector $ |\phi_{n,\xi}(\tau)\rangle $ which is a vector in the Hilbert space of the Bogoliobov mode is defined as
\begin{equation}\label{phic}
|\phi_{n,\xi}(\tau)\rangle=D^{\dagger}(\beta n)S^{\dagger}(\xi)|\beta n e^{-i\tau}, \xi e^{-i2\tau}\rangle.
\end{equation}
Here, the time $ t $ has been replaced with the dimensionless time $ \tau=\Omega_{c}^{\prime}t $. Besides, since the degrees of freedom of the optical field, i.e., operators $ a $ and $ a^{\dagger} $, are not present in these equations and from now on we just deal with the degrees of freedom of the Bogoliubov mode, so we no longer need to retain the index c for identification of the state vectors of the Bogoliubov mode and therefore we have deleted it.

\section{$Q$ function of the Bogoliubov mode}\label{secQ}
Using the reduced density operator of the Bogoliubov mode, one can calculate any of the distribution functions of this mode. Here, we consider the temporal behaviour of the $ Q $ function \cite{Schleich} of this mode and investigate the effect of atom-atom interaction on the shape of this function. The $ Q $ function of the mode $ c $ is defined as
\begin{equation}
Q_{c}(\gamma,\tau)=\frac{1}{\pi}\langle\gamma|\rho_{c}(\tau)|\gamma\rangle,
\end{equation}
where $ \gamma $ is a c-number. Substituting the right hand side of Eq.(\ref{rhoc}) for $ \rho_{c}(\tau) $ in the above equation, the $ Q_{c} $ function reads:
\begin{equation}\label{Q}
Q_{c}(\gamma,\tau)=\frac{1}{\pi}e^{-|\alpha|^2}\sum_{n=0}^{\infty}\frac{|\alpha|^{2n}}{n!}|\langle\gamma|\phi_{n,\xi}(\tau)\rangle|^2.
\end{equation}
The inner product $ \langle\gamma|\phi_{n,\xi}(\tau)\rangle $ can be evaluated as follows:
\begin{eqnarray}
\langle\gamma|\phi_{n,\xi}(\tau)\rangle&=&\langle\gamma|D^{\dagger}(\beta n)S^{\dagger}(\xi)|\beta n e^{-i\tau}, \xi e^{-i2\tau}\rangle,\nonumber\\
&&=e^{i\beta n\gamma_{I}}\langle\beta n+\gamma, \xi|\beta n e^{-i\tau}, \xi e^{-i2\tau}\rangle,
\end{eqnarray}
where $ \gamma_{I} $ is the imaginary part of $ \gamma $. Therefore, the $ Q $ function of the Bogoliubov mode is obtained as
\begin{equation}\label{Qc}
Q_{c}(\gamma,\tau)=\frac{1}{\pi}e^{-|\alpha|^2}\sum_{n=0}^{\infty}\frac{|\alpha|^{2n}}{n!}|\langle\beta n+\gamma, \xi|\beta n e^{-i\tau}, \xi e^{-i2\tau}\rangle|^2.
\end{equation}

In order to see the behaviour of this function in the course of time, we calculate it at the initial time ($ \tau=0 $) and also at $ \tau=\pi/2 $. The $ Q $ function of the Bogoliubov mode at the initial time is obtained as follows
\begin{equation}\label{Qc0}
Q_{c}(\gamma,\tau=0)=\frac{1}{\pi}e^{-|\gamma|^2}.
\end{equation}

To obtain the function $ Q_{c} $ at $ \tau=\pi/2 $ we need to calculate the inner product inside the absolute value in Eq.(\ref{Qc}) which we will denote it by $ f $:
\begin{eqnarray}\label{f}
f&=&\langle\beta n+\gamma, \xi|\beta n e^{-i\frac{\pi}{2}}, \xi e^{-i\pi}\rangle\nonumber\\
&&=\langle\beta n+\gamma, \xi|-i\beta n, -\xi \rangle.
\end{eqnarray}
Using the definition of squeezed states, $ f $ can be written in the following form
\begin{eqnarray}
f&=&\langle\beta n+\gamma|S^{\dagger}(\xi)S(-\xi)|-i\beta n\rangle\nonumber\\
&&=\langle\beta n+\gamma|S(-2\xi)|-i\beta n\rangle\nonumber\\
&&=\langle\beta n+\gamma|-i\beta n,-2\xi\rangle\nonumber\\
&&=\langle\beta n+\gamma|-i\beta n,2\xi e^{i\pi}\rangle.
\end{eqnarray}
The last line in the above equation is the inner product of a squeezed and a coherent state which can be calculated \cite{vogel}. In this way $ |f|^2 $ is obtained as follows
\begin{eqnarray}
|f|^{2}&=&\frac{1}{\mu^{\prime}}\exp\Big[(\frac{\nu^{\prime}}{\mu^{\prime}}-1)\gamma_{R}^{2}-(\frac{\nu^{\prime}}{\mu^{\prime}}+1)\gamma_{I}^{2}+2\beta n(\frac{\nu^{\prime}}{\mu^{\prime}}-1)\gamma_{R}\nonumber\\
&&-2\frac{\beta n}{\mu^{\prime}}\gamma_{I}+2(\frac{\nu^{\prime}}{\mu^{\prime}}-1)\beta^{2}n^2\Big],
\end{eqnarray}
where, by definition, $ \mu^{\prime}=\cosh2\xi $ and $ \nu^{\prime}=\sinh 2\xi $. Substituting this expression for the squared absolute value in Eq.(\ref{Qc}), the function $ Q_{c} $ at the instant $ \tau=\pi/2 $ is obtained as follows
\begin{eqnarray}\label{Qcpi}
Q_{c}(\gamma,\tau&=&\frac{\pi}{2})=\frac{1}{\pi\mu^{\prime}}e^{-|\alpha|^{2}+(\frac{\nu^{\prime}}{\mu^{\prime}}-1)\gamma_{R}^{2}-(\frac{\nu^{\prime}}{\mu^{\prime}}+1)\gamma_{I}^{2}}\nonumber\\
&&\sum_{n=0}^{\infty}\frac{1}{n!}|\alpha|^{2n}e^{2\beta n(\frac{\nu^{\prime}}{\mu^{\prime}}-1)\gamma_{R}-2\frac{\beta n}{\mu^{\prime}}\gamma_{I}+2(\frac{\nu^{\prime}}{\mu^{\prime}}-1)\beta^{2}n^2}.\nonumber\\
\end{eqnarray}

\begin{figure}[ht]
\centering
\includegraphics[width=2.5in]{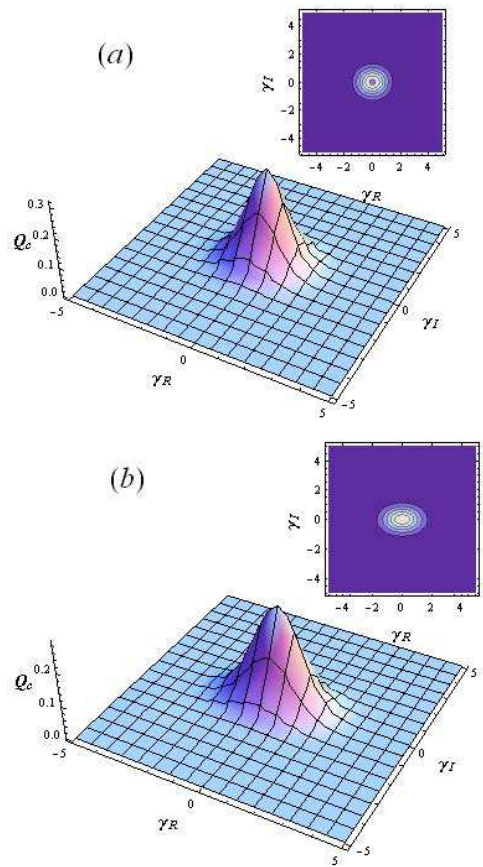}
\caption{
(Color online) The $ Q $ function of the Bogoliubov mode for (a) $ \tau=0 $ and (b) $ \tau=\pi/2 $. The \textit{s}-wave scattering frequency of atomic collisions is $ \omega_{sw}=20\omega_{R} $ and the intensity of the optical field inside the cavity is $ |\alpha|^2=0.01 $. The values of other parameters have been chosen based on the experimental data given in Ref.\cite{Ritter Appl. Phys. B}.}
\label{fig:fig2}
\end{figure}

In order to examine the effect of atom-atom interaction on the behaviour of the $ Q $ function of the Bogoliubov mode we have plotted the function $ Q_{c}(\gamma,\tau) $ in Fig.\ref{fig:fig2}	 at the times $ \tau=0 $ and $ \tau=\pi/2 $ [Eqs.(\ref{Qc0}) and (\ref{Qcpi})] for $ \omega_{sw}=20\omega_{R} $. For this purpose, we have obtained our results based on the experimentally feasible parameters given in Ref.\cite{Ritter Appl. Phys. B} in which the number of $ N=10^5 $ Rubidium atoms distributed in the optical cavity of length $ L=178\mu $m with bare resonance frequency $ \omega_{c} $ corresponding to a wavelength of $ \lambda=780 $nm. The atom-field coupling constant $ g_{0}=2\pi\times14.1 $MHz, the detuning between the pump laser and the atomic transition frequencies $ \Delta_{a}=2\pi\times 58 $GHz, and the scattering length of Rubidium atoms $ a_{s}=5 $nm. Besides, we have assumed that the intensity of the optical field inside the cavity is $ |\alpha|^2=0.01 $.

In Fig.\ref{fig:fig2}(a) the $ Q_{c} $ function has been plotted at the initial time [Eq.(\ref{Qc0})]. As is seen, the shape of this function in the phase space is completely symmetric and its cross section is circular. It is due to the fact that the initial state of the Bogoliubov mode has been considered to be a vacuum state. On the other hand, Fig.\ref{fig:fig2}(b) shows the $ Q_{c} $ function at the time $ \tau=\pi/2 $ [Eq.(\ref{Qcpi})]. As is seen, the cross section of this function has been squeezed along the $ \gamma_{I} $ axsis due to the nonlinear effect of atom-atom interaction.

\section{The effect of atomic collisions on the quadrature squeezing of the Bogoliubov mode}\label{secQuad}
The quadratures of the Bogoliubov mode are defined as the operators $ q=\frac{1}{\sqrt{2}}(c+c^{\dagger}) $ and $ p=\frac{1}{\sqrt{2}i}(c-c^{\dagger}) $ which obey the commutation relation $ [q,p]=i $. If the state of the system is squeezed then the quantum fluctuations in one of these quadratures reduce at the expense of increasing fluctuations in the other one. The degree of squeezing in these quadratures are defined in terms of the squeezing parameters
\begin{subequations}\label{Sqp0}
\begin{eqnarray}
S_{q}(\tau)&=&2\langle(\Delta q)^{2}\rangle-1,\label{Sqp0a}\\
S_{p}(\tau)&=&2\langle(\Delta p)^{2}\rangle-1\label{Sqp0b},
\end{eqnarray}
\end{subequations}
where $ \langle(\Delta q)^{2}\rangle=\langle q^2\rangle-\langle q\rangle^2 $ and $ \langle(\Delta p)^{2}\rangle=\langle p^2\rangle-\langle p\rangle^2 $ are the quantum uncertainties. When $ S_{i}(\tau)<0  (i=p, q)$, the corresponding state is a squeezed one.

Using the definition of quadratures and their uncertainties, the squeezed parameters [Eqs.(\ref{Sqp0a}, \ref{Sqp0b})] can be written in terms of the creation and annihilation operators of the Bogoliubov mode as follows:
\begin{subequations}\label{Sqp}
\begin{eqnarray}
S_{q}(\tau)&=&\Big[\langle c^2\rangle+\langle c^{\dagger2}\rangle-\langle c\rangle^2-\langle c^{\dagger}\rangle^2\Big]\label{Sqpa}\nonumber\\
&&+2\Big[\langle c^{\dagger}c\rangle-\langle c\rangle\langle c^{\dagger}\rangle\Big],\\
S_{p}(\tau)&=&-\Big[\langle c^2\rangle+\langle c^{\dagger2}\rangle-\langle c\rangle^2-\langle c^{\dagger}\rangle^2\Big]\label{Sqpb}\nonumber\\
&&+2\Big[\langle c^{\dagger}c\rangle-\langle c\rangle\langle c^{\dagger}\rangle\Big].
\end{eqnarray}
\end{subequations}

\subsection{Calculation of $ \langle c\rangle $ and $  \langle c^{\dagger}\rangle $ }
Using the reduced density operator of the Bogoliubov mode [Eq.(\ref{rhoc})], the expectation value of the annihilation operator can be calculated by the relation $ \langle c\rangle=tr[\rho_{c}c] $. The expectation value of the creation operator can be obtained by the relation $ \langle c^{\dagger}\rangle= \langle c\rangle^{\ast} $.
\begin{eqnarray}\label{mc}
\langle c\rangle&=&tr\Big[e^{-|\alpha|^2}\sum_{n=0}^{\infty}\frac{|\alpha|^{2n}}{n!}|\phi_{n,\xi}(\tau)\rangle\langle\phi_{n,\xi}(\tau)|c\Big]\nonumber\\
&&=e^{-|\alpha|^2}\sum_{n=0}^{\infty}\frac{|\alpha|^{2n}}{n!}\langle\phi_{n,\xi}(\tau)|c|\phi_{n,\xi}(\tau)\rangle
\end{eqnarray}
In order to calculate this summation we first need to evaluate $ \langle c\rangle_{\phi}=\langle\phi_{n,\xi}(\tau)|c|\phi_{n,\xi}(\tau)\rangle $. By using Eq.(\ref{phic}) for $ |\phi_{n,\xi}(\tau)\rangle $ one gets
\begin{eqnarray}\label{picpi}
\langle c\rangle_{\phi}&=&\langle\ s |S(\xi)D(\beta n) c D^{\dagger}(\beta n)S^{\dagger}(\xi)|s\rangle\nonumber\\
&&=\langle\ s|S(\xi) c S^{\dagger}(\xi)-\beta n|s\rangle\nonumber\\
&&=\langle s|\mu c+\nu c^{\dagger}-\beta n|s\rangle\nonumber\\
&&=\mu \langle c\rangle_{s}+\nu \langle c^{\dagger}\rangle_{s}-\beta n,
\end{eqnarray}
where we have represented $ |\beta n e^{-i\tau}, \xi e^{-i2\tau}\rangle $ by the state vector $ |s\rangle $. In the second line we have used Eqs.(\ref{DcDa}, \ref{DcDb}) with the substitution $ n $ for $ a^{\dagger}a $ and in the third line we have made use of Eqs.(\ref{ScSa}, \ref{ScSb}). Now we should calculate $ \langle c\rangle_{s} $:
\begin{eqnarray}
\langle c\rangle_{s}&=&\langle\beta n e^{-i\tau}, \xi e^{-i2\tau}|c |\beta n e^{-i\tau}, \xi e^{-i2\tau}\rangle\nonumber\\
&&=\langle\beta n e^{-i\tau}|S^{\dagger}(\xi e^{-2i\tau}) c S(\xi e^{-2i\tau})|\beta n e^{-i\tau}\rangle\nonumber\\
&&=\langle\beta n e^{-i\tau}|\mu c-\nu e^{-2i\tau}c^{\dagger}|\beta n e^{-i\tau}\rangle.
\end{eqnarray}
In the third line we have used the unitary transformation $ S^{\dagger}(r e^{i\theta})c S(r e^{i\theta})=\mu c-\nu e^{i\theta}c^{\dagger} $. Now, using the eigenvalue equation $ c|\beta n e^{-i\tau}\rangle=\beta n e^{-i\tau}|\beta n e^{-i\tau}\rangle\ $, we will have:
\begin{equation}\label{mcs}
\langle c\rangle_{s}=\beta n(\mu-\nu) e^{-i\tau}=\langle c^{\dagger}\rangle^{\ast}_{s}.
\end{equation}

Using Eqs.(\ref{mcs}) and (\ref{picpi}) the expectation values $  \langle c\rangle $ and $ \langle c^{\dagger}\rangle $ are obtained as follows
\begin{subequations}
\begin{eqnarray}
 \langle c\rangle &=&\beta |\alpha|^2\Big[(\mu-\nu)(\mu e^{-i\tau}+\nu e^{i\tau})-1\Big],\\
 \langle c^{\dagger}\rangle &=&\beta |\alpha|^2\Big[(\mu-\nu)(\mu e^{i\tau}+\nu e^{-i\tau})-1\Big].
\end{eqnarray}
\end{subequations}

\subsection{Calculation of $ \langle c^2\rangle $ and $ \langle c^{\dagger2}\rangle $}

The expectation value $ \langle c^2\rangle $ is calculated from the relation $ \langle c^2\rangle=tr[\rho_{c}c^2] $ and then $ \langle c^{\dagger2}\rangle $ will be obtained from its complex conjugate, i.e., $ \langle c^{\dagger2}\rangle=\langle c^2\rangle^{\ast} $. In this way, using the reduced density operator [Eq.(\ref{rhoc})] the expectation value $ \langle c^2\rangle $ is calculated in the following way:
\begin{eqnarray}\label{mc2s1}
\langle c^2\rangle&=&tr\Big[e^{-|\alpha|^2}\sum_{n=0}^{\infty}\frac{|\alpha|^{2n}}{n!}|\phi_{n,\xi}(\tau)\rangle\langle\phi_{n,\xi}(\tau)|c^2\Big]\nonumber\\
&&=e^{-|\alpha|^2}\sum_{n=0}^{\infty}\frac{|\alpha|^{2n}}{n!}\langle\phi_{n,\xi}(\tau)|c^2|\phi_{n,\xi}(\tau)\rangle.
\end{eqnarray}
Now, using the definition of the state vector $ |\phi_{n,\xi}(\tau)\rangle $ given in Eq.(\ref{phic}) the averaged value appeared in the above summation, i.e., $ \langle c^2\rangle_{\phi}= \langle\phi_{n,\xi}(\tau)|c^2|\phi_{n,\xi}(\tau)\rangle$ can be calculated from the following relation
\begin{equation}\label{mc2phi}
\langle c^2\rangle_{\phi}=\langle\beta n e^{-i\tau}, \xi e^{-i2\tau}|X c^{2} X^{\dagger}|\beta n e^{-i\tau}, \xi e^{-i2\tau}\rangle,
\end{equation}
where the operator $ X=S(\xi)D(\beta n) $ . Using the unitary property of this operator one can substitute $ (XcX^{\dagger})^2  $ for $ Xc^2X^{\dagger} $ in the above relation. Therefore, we should first calculate $ XcX^{\dagger} $ in the following way
\begin{eqnarray}\label{xcxdag}
XcX^{\dagger}&=&S(\xi)D(\beta n) c D^{\dagger}(\beta n)S^{\dagger}(\xi)\nonumber\\
&&=S(\xi) (c-\beta n) S^{\dagger}(\xi)\nonumber\\
&&=\mu c+ \nu c^{\dagger}-\beta n.
\end{eqnarray}
In the second and third lines we have used respectively, the transformations (\ref{DcD}) and (\ref{ScS}). Now, substituting the square of the right hand side of the above equation for $ Xc^{2}X $ in Eq.(\ref{mc2phi}) and using the definition $ |s\rangle=|\beta n e^{-i\tau}, \xi  e^{-2i\tau}\rangle $ we will have
\begin{eqnarray}\label{mc2phiR}
\langle c^2\rangle_{\phi}&=&\langle s|(XcX^{\dagger})^2|s\rangle\nonumber\\
&&=\langle s|(\mu c+ \nu c^{\dagger}-\beta n)^2|s\rangle\nonumber\\
&&=\mu^2\langle c^2\rangle_{s}+\nu^2\langle c^{\dagger2}\rangle_{s}+\mu\nu+2\mu\nu\langle c^{\dagger}c\rangle_{s}\nonumber\\
&&-2\mu\beta n\langle c\rangle_{s}-2\nu\beta n\langle c^{\dagger}\rangle_{s}+\beta^2n^2.
\end{eqnarray}

In order to obtain the right hand side of Eq.(\ref{mc2phiR}) we need to have all the averaged values in the state $ |s\rangle $. We have already obtained the averaged values $ \langle c\rangle_{s} $ and $ \langle c^{\dagger}\rangle_{s} $. Now, we should calculate $  \langle c^2\rangle_{s} $ and $ \langle c^{\dagger}c\rangle_{s} $. For the former we act as follows
\begin{eqnarray}
\langle c^2\rangle_{s}&=&\langle\beta n e^{-i\tau}, \xi e^{-i2\tau}|c^2 |\beta n e^{-i\tau}, \xi e^{-i2\tau}\rangle\nonumber\\
&&=\langle\beta n e^{-i\tau}|S^{\dagger}(\xi e^{-2i\tau}) c^2 S(\xi e^{-2i\tau})|\beta n e^{-i\tau}\rangle\nonumber\\
&&=\langle\beta n e^{-i\tau}|\Big(S^{\dagger}(\xi e^{-2i\tau}) c S(\xi e^{-2i\tau})\Big)^2|\beta n e^{-i\tau}\rangle\nonumber\\
&&=\langle\beta n e^{-i\tau}|(\mu c-\nu e^{-2i\tau}c^{\dagger})^2|\beta n e^{-i\tau}\rangle.
\end{eqnarray}
Here, all the steps of calculation are similar to those done for $ \langle c\rangle_{s} $. By expanding the squared experssion in last line and noting that $ |\beta n e^{-i\tau}\rangle $ is the eigenvector of the operator $ c $ the following result is obtained:
\begin{equation}\label{mc2s}
\langle c^2\rangle_{s}=\Big[\beta^2 n^2(\mu-\nu)^2-\mu\nu\Big] e^{-2i\tau}.
\end{equation}
On the other hand, to obtain $ \langle c^{\dagger}c\rangle_{s} $ we have
\begin{eqnarray}
\langle c^{\dagger}c\rangle_{s}&=&\langle\beta n e^{-i\tau}, \xi e^{-i2\tau}|c^{\dagger}c |\beta n e^{-i\tau}, \xi e^{-i2\tau}\rangle\nonumber\\
&&=\langle\beta n e^{-i\tau}|S^{\dagger}(\xi e^{-2i\tau}) c^{\dagger}c S(\xi e^{-2i\tau})|\beta n e^{-i\tau}\rangle\nonumber\\
&&=\langle\beta n e^{-i\tau}|S^{\dagger}c^{\dagger}S S^{\dagger}c S)|\beta n e^{-i\tau}\rangle\nonumber\\
&&=\langle\beta n e^{-i\tau}|(\mu c^{\dagger}-\nu e^{2i\tau}c)\nonumber\\
&& \times (\mu c-\nu e^{-2i\tau}c^{\dagger})|\beta n e^{-i\tau}\rangle.
\end{eqnarray}
Then, by expanding the expression in the last two lines and using the eigenvalue equations for $ c $ and $ c^{\dagger} $ the expectation value $ \langle c^{\dagger}c\rangle_{s} $ can be obtained as follows
\begin{equation}
\langle c^{\dagger}c\rangle_{s}=\beta^2n^2(\mu-\nu)^2+\nu^2.
\end{equation}

Now, substituting the derived expectation values in the state $ |s\rangle $ in Eq.(\ref{mc2phiR}), $ \langle c^2\rangle_{\phi} $ is obtained as follows
\begin{equation}\label{mcpfif}
\langle\phi_{n,\xi}(\tau)|c^2|\phi_{n,\xi}(\tau)\rangle=\beta^2n^2f_{1}+f_{2},
\end{equation}
where $ f_{1} $ and $ f_{2} $ have been defined as
\begin{subequations}
\begin{eqnarray}
f_{1}&=&(\mu-\nu)^2(\mu^2e^{-2i\tau}+\nu^2e^{2i\tau})+2\mu\nu(\mu-\nu)^2\nonumber\\
&&-2(\mu-\nu)(\mu e^{-i\tau}+\nu e^{i\tau})+1,\label{f1a}\\
f_{2}&=&-\mu\nu(\mu^2e^{-2i\tau}+\nu^2e^{2i\tau})+\mu\nu+2\mu\nu^3\label{f2b}.
\end{eqnarray}
\end{subequations}

In this way the expectation value $ \langle c^2\rangle $ and its complex conjugate are derived as follows
\begin{subequations}
\begin{eqnarray}
\langle c^2\rangle &=& |\alpha |^2(1+|\alpha |^2)\beta^2f_{1}+f_{2},\label{c2a}\\
\langle c^{\dagger2}\rangle &=& |\alpha |^2(1+|\alpha |^2)\beta^2f_{1}^{\ast}+f_{2}^{\ast}\label{c2b}.
\end{eqnarray}
\end{subequations}

\subsection{Calculation of $ \langle c^{\dagger}c\rangle $}
In the final step we need to calculate $ \langle c^{\dagger}c\rangle $. This expectation value is obtained from the relation $ \langle c^{\dagger}c\rangle=tr[\rho_{c}c^{\dagger}c] $ in the following form:
\begin{eqnarray}\label{mcdagcs1}
\langle c^{\dagger}c\rangle&=&tr\Big[e^{-|\alpha|^2}\sum_{n=0}^{\infty}\frac{|\alpha|^{2n}}{n!}|\phi_{n,\xi}(\tau)\rangle\langle\phi_{n,\xi}(\tau)|c^{\dagger}c\Big]\nonumber\\
&&=e^{-|\alpha|^2}\sum_{n=0}^{\infty}\frac{|\alpha|^{2n}}{n!}\langle\phi_{n,\xi}(\tau)|c^{\dagger}c|\phi_{n,\xi}(\tau)\rangle.
\end{eqnarray}
Using the definition of the state vector $ |\phi_{n,\xi}(\tau)\rangle $ [Eq.(\ref{phic})] the expectation value $ \langle c^{\dagger}c\rangle_{\phi}=\langle\phi_{n,\xi}(\tau)|c^{\dagger}c|\phi_{n,\xi}(\tau)\rangle $ can be calculated:
\begin{equation}\label{mcdagcphis1}
\langle c^{\dagger}c\rangle_{\phi}=\langle\beta n e^{-i\tau}, \xi e^{-i2\tau}|X c^{\dagger}c X^{\dagger}|\beta n e^{-i\tau}, \xi e^{-i2\tau}\rangle,
\end{equation}
where as before the operator $ X $ has been defined as $ X=S(\xi)D(\beta n) $. By using the unitarity of the operator $ X $ and applying Eq.(\ref{xcxdag}) we get
\begin{eqnarray}
Xc^{\dagger}cX^{\dagger}&=&Xc^{\dagger}X^{\dagger}XcX^{\dagger}\nonumber\\
&&=(\nu c+\mu c^{\dagger}-\beta n)(\mu c+\nu c^{\dagger}-\beta n),\nonumber\\
&&=\mu\nu(c^2+c^{\dagger2})+(\mu^2+\nu^2)c^{\dagger}c\nonumber\\
&&-\beta n(\mu+\nu)(c+c^{\dagger})+\beta^2n^2+\nu^2.
\end{eqnarray}
Substituting this expression into Eq.(\ref{mcdagcphis1}) we will have
\begin{eqnarray}
\langle c^{\dagger}c\rangle_{\phi}&=&\mu\nu(\langle c^2\rangle_{s}+\langle c^{\dagger2}\rangle_{s})+(\mu^2+\nu^2)\langle c^{\dagger}c\rangle_{s}\nonumber\\
&&-\beta n(\mu+\nu)(\langle c\rangle_{s}+\langle c^{\dagger}\rangle_{s})+\beta^2n^2+\nu^2.
\end{eqnarray}
Now, using the the averaged values $ \langle c\rangle_{s}, \langle c^2\rangle_{s} $ and their complex conjugate that have been already calculated, we can obtain $ \langle c^{\dagger}c\rangle_{\phi} $ in the following form
\begin{equation}\label{mcdagcphi}
\langle\phi_{n,\xi}(\tau)|c^{\dagger}c|\phi_{n,\xi}(\tau)\rangle=\beta^2n^2 f_{3}+f_{4},
\end{equation}
where $ f_{3} $ and $ f_{4} $ have been defined as follows
\begin{subequations}
\begin{eqnarray}
f_{3}&=&2\mu\nu(\mu-\nu)^2\cos2\tau+(\mu^2+\nu^2)(\mu-\nu)^2\nonumber\\
&&-2\cos\tau+1,\\
f_{4}&=&-2\mu^2\nu^2\cos2\tau+\nu^2(\mu^2+\nu^2)+\nu^2.
\end{eqnarray}
\end{subequations}
By substituting Eq.(\ref{mcdagcphi}) into Eq.(\ref{mcdagcs1}) we finally arrive at
\begin{equation}\label{mcdagc}
\langle c^{\dagger}c\rangle=\beta^2f_{3}|\alpha|^2(1+|\alpha|^2)+f_{4}.
\end{equation}

\subsection{Dynamical behaviour of the squeezing parameters for the Bogoliubov mode}

The squeezing parameters have been obtained in terms of the averaged values of the operators $ c $ and $ c^{\dagger} $ in Eqs.(\ref{Sqpa}, \ref{Sqpb}). Now using the expressions obtained for these averaged valued the squeezing parameters are obtained as follows:
\begin{subequations}\label{spqf}
\begin{eqnarray}
S_{q}(\tau)&=&4\beta^2|\alpha|^2(1-\cos\tau)^2\nonumber\\
&&+2\nu(\mu+\nu)\Big[1-(\mu+\nu)(\mu\cos2\tau-\nu)\Big],\label{spqfa}\\
S_{p}(\tau)&=&4\beta^2|\alpha|^2(\mu-\nu)^4\sin^{2}(\tau)\nonumber\\
&&+2\nu(\mu-\nu)\Big[(\mu-\nu)(\mu\cos2\tau+\nu)-1\Big].\label{spqfb}
\end{eqnarray}
\end{subequations}

These are functions of $ \mu $ and $ \nu $ which depend on the nonlinearity parameter $ \omega_{sw} $ through Eqs.(\ref{munua}, \ref{munub}) and (\ref{Omepc}). If the nonlinearity parameter is zero ($ \omega_{sw}=0 $) then based on Eq.(\ref{Omepc}) we have $ \Omega^{\prime}_{c}=\Omega_{c} $. Substituting this in Eqs.(\ref{munua}, \ref{munub}) the values of $ \mu $ and $ \nu $ are obtained as $ \mu=1 $ and $ \nu=0 $. In this way, in the absence of the nonlinear effect of atom-atom interaction the squeezing parameters are obtained in the following forms:
\begin{subequations}\label{sw0}
\begin{eqnarray}
S_{p}(\tau)|_{\omega_{sw}=0}&=&4\beta^2|\alpha|^2\sin^{2}\tau,\label{sw0a}\\
S_{q}(\tau)|_{\omega_{sw}=0}&=&4\beta^2|\alpha|^2(1-\cos\tau)^2\label{sw0b}.
\end{eqnarray}
\end{subequations}

These equations show clearly that in the absence of the nonlinear effects of atomic collisions the values of the both squeezing parameters are always positive. In the other words, when the nonlinearity parameter is zero the squeezing property of the Bogoliubov mode disappears.

\begin{figure}[ht]
\centering
\includegraphics[width=2.7in]{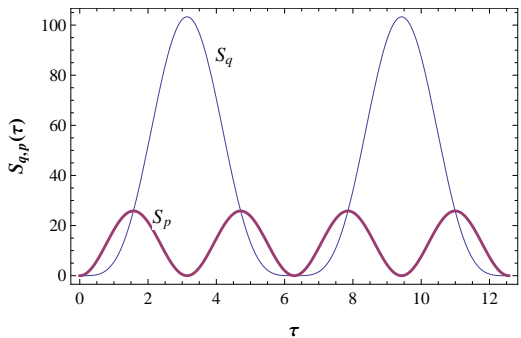}
\caption{
(Color online) The squeezing parameters $ S_{q} $ (blue line) and $ S_{p} $ (purple line) versus the dimensionless time $ \tau $ for $ \omega_{sw}=0 $. The values of the other parameters have been chosen based on the experimental data of Ref.\cite{Ritter Appl. Phys. B} just like those of Fig.\ref{fig:fig2}}
\label{fig:fig3}
\end{figure}

In Fig.(\ref{fig:fig3}) the squeezing parameters corresponding to the quadratures of the Bogoliubov mode in the absence of the nonlinear effect of atomic collisions [Eqs(\ref{sw0a}, \ref{sw0b})] have been plotted versus the dimensionless time $ \tau $. The values of the parameters in this figure, just like those of Fig.(\ref{fig:fig2}), have been chosen based on the experimental data of Ref.\cite{Ritter Appl. Phys. B}.

\begin{figure}[ht]
\centering
\includegraphics[width=2.7in]{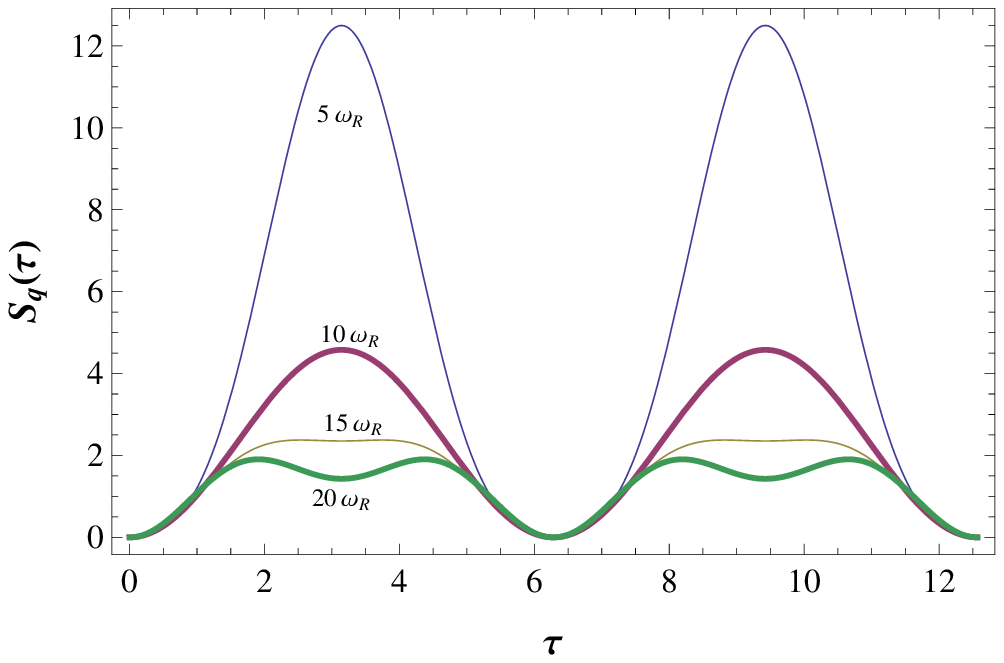}
\includegraphics[width=2.7in]{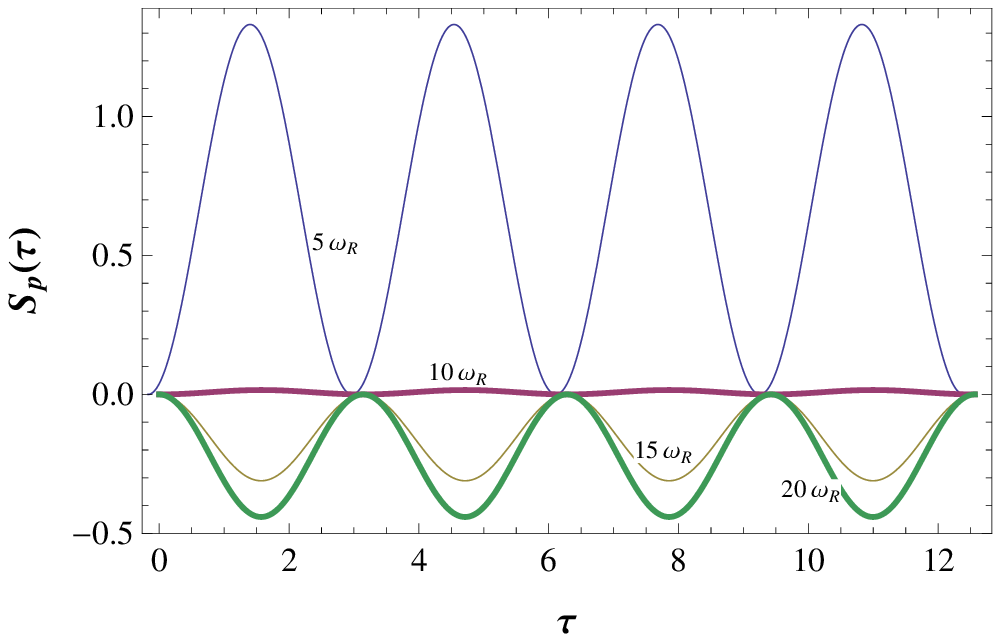}
\caption{
(Color online) (a) the squeezing parameter $ S_{q} $ and (b) the squeezing parameter $ S_{p} $ versus the dimensionless time $ \tau $ for different values of the \textit{s}-wave scattering frequency. The values of the other parameters have been chosen based on the experimental data of Ref.\cite{Ritter Appl. Phys. B} just like those of Fig.\ref{fig:fig2}}.
\label{fig:fig4}
\end{figure}

On the other hand, considering the nonlinear effect of atom-atom interaction, i.e., when $ \omega_{sw}\neq0 $, the parameters $ \mu $ and $ \nu $ are both nonzero and based on Eqs.(\ref{spqfa}, \ref{spqfb}) the squeezing property appears in the Bogoliubov mode. In Fig.\ref{fig:fig4} the squeezing parameters have been plotted versus the dimensionless time $ \tau $ for different nonzero values of the \textit{s}-wave scattering parameter. In Fig.\ref{fig:fig4}(a) the parameter $ S_{q}(\tau) $ [Eq.(\ref{spqfa})] and in Fg.\ref{fig:fig4}(b) the parameter $ S_{p}(\tau) $ [Eq.(\ref{spqfb})] have been plotted for $ \omega_{sw}=5\omega_{R} $ (blue line), $ \omega_{sw}=10\omega_{R} $ (purple line), $ \omega_{sw}=15\omega_{R} $ (brown line) and $ \omega_{sw}=20\omega_{R} $ (green line).

As is seen from these figures, $ S_{q}(\tau) $ is always positive for all values of the \textit{s}-wave scattering frequency. It means that the squeezing property does not appear in the quadrature $ q $. Instead, the squeezing parameter $ S_{p}(\tau) $ for $ \omega_{sw}> 10\omega_{R} $ is always negative which means that the increase of the nonlinearity parameter (the \textit{s}-wave scattering frequency of atomic collisions) causes the squeezing property to appear in the quadrature $ p $.

\section{Conclusion}\label{secCon}
In conclusion, we we have introduced a physical scheme that allows one to generate and control the nonclassical squeezed states in the Bogoliubov mode of a one-dimensional Bose-Einstein condensate trapped inside the optical lattice of a cavity which interacts dispersively with the optical field of the cavity. 

In the weak-interaction regime and in the Bogoliubov approximation, the atomic field of the BEC can be described by a single mode quantum field in a simple optomechanical model in which the quantum fluctuations of the atomic field (Bogoliubov mode) is coupled to the radiation pressure of the optical field of the cavity.

Using the similarity between the atomic interaction in the present model and the interaction potential of a DPA, we have shown that the condensate and the Bogoliubov modes in this hybrid system play, respectively, the roles of the pump field and the signal mode in the DPA system. In fact, as the nonlinear property of the crystal in the DPA system causes the manifestation of quadrature squeezing in the signal mode, in the same way the nonlinear effect of atom-atom interaction in the BEC can lead to the generation of squeezing property in the Bogoliubove mode of the BEC.

Therefore, the hybrid system under consideration behaves just like a parametric amplifier in which the \textit{s}-wave scattering frequency plays the role of the nonlinear gain parameter. Since the \textit{s}-wave scattering frequency is controllable by the transverse harmonic potential confining the BEC, so the degree of squeezing of the Bogoliubov mode can be effectively controlled.

\section*{Acknowledgement}
The authors wish to thank The Office of Graduate Studies of The University of Isfahan for its support.

\appendix
\section{The free evolution of a squeezed coherent state}\label{apA}
In this appendix we derive the free evolution of a squeezed sate \cite{stoler, ma}. Let us consider a free field described by the Hamiltonian $ H=\hbar\omega_{c}c^{\dagger}c $ and assume that the system has been prepared in an initial squeezed coherent state $ |\beta, \xi\rangle=S(\xi)D(\beta)|0\rangle $ where $ \xi=re^{i\theta} $ is the squeezing parameter, the state $ |\beta\rangle $ is the eigenstate of the annihilation operator $ c $, i.e., $ c|\beta\rangle=\beta|\beta\rangle $, and $ |0\rangle $ is the vacuum state of the field. The displacement and squeezing operators is defined as
\begin{subequations}
\begin{eqnarray}
D(\beta)&=&e^{\beta c^{\dagger}-\beta^{\star}c},\\
S(\xi)&=&e^{\frac{1}{2}(\xi^{\star}c^2-\xi c^{\dagger2})},
\end{eqnarray}
\end{subequations}
where $ \xi $ and $ \beta $ are c-numbers. We are going to obtain the state vector of the system at time $ t $ which can be written as follows
\begin{equation}\label{sqt}
|\beta,\xi\rangle_{t}=e^{-i\omega_{c}t c^{\dagger}c}|\beta,\xi\rangle.
\end{equation}
For this purpose we should first consider the transformation of the operators $ c $ and $ c^{\dagger} $ under the unitary operator $ S(\xi) $ which are given in the following form:
\begin{subequations}
\begin{eqnarray}
c^{\prime}&=&S(\xi) c S^{\dagger}(\xi)= \mu c+\nu e^{i\theta} c^{\dagger},\label{cpa}\\
c^{\prime\dagger}&=&S(\xi) c^{\dagger} S^{\dagger}(\xi)= \nu e^{-i\theta} c+\mu c^{\dagger}.
\end{eqnarray}
\end{subequations}

Noting the definition of a squeezed coherent state as $ |\beta, \xi\rangle=S(\xi)|\beta\rangle $ we can derive the action of the transformed operator $ c^{\prime} $ on the state $ |\beta, \xi\rangle $ as
\begin{eqnarray}
c^{\prime}|\beta, \xi\rangle&=&S(\xi) c S^{\dagger}(\xi) S(\xi) |\beta\rangle=S(\xi) c |\beta\rangle\nonumber\\
&&=\beta S(\xi)|\beta\rangle=\beta |\beta,\xi\rangle.
\end{eqnarray}
Substituting for $ c^{\prime} $ from Eq.(\ref{cpa}) in the above equation we obtain:
\begin{equation}\label{egsq}
(\mu c+\nu e^{i\theta} c^{\dagger})|\beta,\xi\rangle=\beta |\beta,\xi\rangle.
\end{equation}
Based on this equation $ |\beta,\xi\rangle $ is an eigenstate of the annihilation operator $ c^{\prime}=\mu c+\nu e^{i\theta} c^{\dagger} $ with the eigenvalue $ \beta $. Now we want to show that $ |\beta,\xi\rangle_{t} $ is the eigenstate of the operator $ C\equiv\mu e^{i\omega_{c}t} c+\nu e^{i\theta} e^{-i\omega_{c}t} c^{\dagger} $. For this purpose we should determine the action of this operator on the mentioned state vector:
\begin{eqnarray}\label{C}
C|\beta,\xi\rangle_{t}&=&C e^{-i\omega_{c}t c^{\dagger}c}|\beta,\xi\rangle\nonumber\\
&&=e^{-i\omega_{c}t c^{\dagger}c}e^{+i\omega_{c}t c^{\dagger}c} C e^{-i\omega_{c}t c^{\dagger}c}|\beta,\xi\rangle.
\end{eqnarray}
Now noting that
\begin{subequations}
\begin{eqnarray}
e^{+i\omega_{c}t c^{\dagger}c} c e^{-i\omega_{c}t c^{\dagger}c}&=&ce^{-i\omega_{c}t},\\
e^{+i\omega_{c}t c^{\dagger}c} c^{\dagger} e^{-i\omega_{c}t c^{\dagger}c}&=&c^{\dagger}e^{+i\omega_{c}t},
\end{eqnarray}\label{PeyA}
\end{subequations}
we will have
\begin{equation}
e^{+i\omega_{c}t c^{\dagger}c} C e^{-i\omega_{c}t c^{\dagger}c}=(\mu c+\nu e^{i\theta} c^{\dagger}).
\end{equation}
Substituting this result in Eq.(\ref{C}) we will have
\begin{eqnarray}\label{egsqt}
C|\beta,\xi\rangle_{t}&=&e^{-i\omega_{c}t c^{\dagger}c}(\mu c+\nu e^{i\theta} c^{\dagger})|\beta,\xi\rangle\nonumber\\
&&=\beta e^{-i\omega_{c}t c^{\dagger}c}|\beta,\xi\rangle\nonumber\\
&&=\beta |\beta,\xi\rangle_{t}.
\end{eqnarray}
As is seen from Eq.(\ref{egsqt}), $ |\beta,\xi\rangle_{t} $ is the eigenstate of the operator $ C $ with the eigenvalue $ \beta $. Now, multiplying both sides of Eq.(\ref{egsqt}) by $ e^{-i\omega_{c}t} $ we will have
\begin{eqnarray}
e^{-i\omega_{c}t}C|\beta,\xi\rangle_{t}&=&e^{-i\omega_{c}t}(\mu e^{i\omega_{c}t} c+\nu e^{i\theta} e^{-i\omega_{c}t} c^{\dagger})|\beta,\xi\rangle_{t}\nonumber\\
&&=(\mu c+\nu e^{i(\theta-2\omega_{c}t)} c^{\dagger})|\beta,\xi\rangle_{t}\nonumber\\
&&=\beta e^{-i\omega_{c}t}|\beta,\xi\rangle_{t}.
\end{eqnarray}

On the other hand comparing this equation with Eq.(\ref{egsq}) which can be rewritten in the following form
\begin{equation}\label{sqtn}
(\mu c+\nu e^{i\theta} c^{\dagger})|\beta,r e^{i\theta}\rangle=\beta |\beta,r e^{i\theta}\rangle,
\end{equation}
one can conclude that $ |\beta,\xi\rangle_{t} $ is a squeezed coherent state:
\begin{equation}\label{sqtresult}
|\beta,\xi\rangle_{t}=|\beta e^{-i\omega_{c}t}, \xi e^{-i2\omega_{c}t}\rangle.
\end{equation}
In this way considering Eqs.(\ref{sqt}) and (\ref{sqtresult}) one goes to the conclusion that the free evolution of an initial squeezed coherent state is given as follows
\begin{equation}\label{fesq}
e^{-i\omega_{c}t c^{\dagger}c}|\beta,\xi\rangle=|\beta e^{-i\omega_{c}t}, \xi e^{-i2\omega_{c}t}\rangle.
\end{equation}

\bibliographystyle{apsrev4-1}

\end{document}